\documentclass[
  reprint,
amsmath,amssymb,
aps,
]{revtex4-2}
\usepackage{graphicx}
\usepackage{color}
\usepackage{dcolumn}
\usepackage{bm}
\usepackage{hyperref}

\hypersetup{colorlinks,linkcolor={blue},citecolor={blue},urlcolor={blue}}
\usepackage{float}

\usepackage{cancel}
\usepackage[normalem]{ulem}
\usepackage{xcolor}
\usepackage{color}
\newcommand{\beq}{\begin{equation}}
\newcommand{\eeq}{\end{equation}}
\def\barr{\begin{array}}
\def\earr{\end{array}}
\def\dis{\displaystyle}
\definecolor{darkcyan}{cmyk}{1,0,0,0.4}
\definecolor{darkgreen}{cmyk}{1,0,1,0.4}

\def\lapp{\mathrel{\rlap{\raise.5ex\hbox{$<$}}
                    {\lower.5ex\hbox{$\sim$}}}}
\def\gapp{\mathrel{\rlap{\raise.5ex\hbox{$>$}}
                    {\lower.5ex\hbox{$\sim$}}}}

\def\d{{\rm d}}
\def\x{{\rm \bf x}}
\def\k{{\rm \bf k}}

\def\pa{{\partial}}

\def\beq{\begin{equation}}
	\def\eeq{\end{equation}}
\def\bea{\begin{eqnarray}}
	\def\eea{\end{eqnarray}}

\begin{document}
	
	\preprint{APS/123-QED}
	
	\title{Dark photon---Assisted Primordial Magnetogenesis}
	
	\author{Debottam Nandi${}^1$}
	\email{debottam.nandi@vit.ac.in}

	\author{Debajyoti Choudhury${}^2$}%
	\email{debchou@physics.du.ac.in}
	\affiliation{${}^1$Department of Physics, School of Advanced Sciences, Vellore Institute of Technology (VIT) Chennai, Chennai 600127, India}

	\affiliation{${}^2$Department of Physics and Astrophysics, University of Delhi, Delhi 110007, India}
	
\begin{abstract}
Magnetic fields observed across cosmic scales are difficult to explain
within conventional physics. A primordial origin is, thus, often
assumed. While a nonminimal coupling of the inflaton with the
electromagnetic field could theoretically generate magnetic fields of
about $10^{-13}$ G, this approach faces significant issues, including
strong-coupling and backreaction problems. ``Dark photons", arising
naturally in hidden-sector extensions of the Standard Model, provide a
well-motivated framework for addressing various cosmic as well as
particle physics issues. We demonstrate that coupling
dark photons with standard ones can result in adequate
magnetogenesis without the limitations of existing models. This
minimal mechanism may also provide insights into unresolved cosmic
mysteries.
\end{abstract}
	
	\maketitle

\section{Introduction}

Large scale coherent magnetic fields are ubiquitous in the Universe
across scales, from galaxies and clusters to intergalactic voids, with
strengths ranging from $\mu{\rm G}$ down to
$10^{-15}$--$10^{-18}\,{\rm G}$. The coherence lengths strongly
suggest a primordial origin, motivating extensive studies of cosmic
magnetogenesis in the early Universe
\cite{Kandus:2010nw,Durrer:2013pga,Subramanian:2015lua,Vachaspati:2020blt}.
While it is appealing to ascribe correlations on super-Hubble scales
to an inflationary origin, the conformal invariance of classical
electrodynamics coupled minimally to gravity prevents magnetic field
amplification during inflation
\cite{Turner:1987bw,Ratra:1991bn}. Although breaking the 
invariance could lead to significant abundance of magnetic field, such
models generically encounter severe theoretical difficulties. In
particular, efficient amplification typically implies either an
unadmissibly large effective gauge coupling (the strong-coupling
problem) or a significant backreaction of the generated
electromagnetic fields on the inflationary background
\cite{Demozzi:2009fu}.
Reexaminations~\cite{Fujita:2012rb,Kobayashi:2014sga} led to robust
bounds on inflationary magnetogenesis from such considerations.

To evade these, several extensions have been explored, including
axion-like couplings, spectator fields, and alternative time profiles
of the gauge kinetic function
\cite{Anber:2006xt,Ferreira:2013sqa,Nandi:2021lpf,Nandi:2017ajk}. An alternative is magnetogenesis in the
presence of multiple gauge fields. Models with two interacting $U(1)$
sectors naturally arise in extensions of the Standard Model and allow
for energy transfer between visible and hidden sectors
\cite{Giovannini:1999wv,BeltranJimenez:2010brg}.  Such dark photons have been
studied extensively in cosmology and particle physics, with
applications ranging from dark radiation to viable dark matter
candidates \cite{Holdom:1985ag,Redondo:2010dp,Arias:2012az}.  These could
participate non-trivially in inflationary dynamics while remaining
consistent with laboratory, astrophysical, and cosmological
constraints, making them natural ingredients in scenarios that seek to
transfer energy between visible and dark sectors.

Despite their appeal, implementing dark photons in inflationary
magnetogenesis have been claimed~\cite{Ferreira:2014mwa} to face the
aforementioned tension.  We argue that such a conclusion 
is based on very strong and non-generic
assumptions. Allowing, instead, for a controlled and transient
interaction between the photon and a dark photon provides a minimal
and robust realization of inflationary magnetogenesis. The resulting
amplification avoids both strong-coupling and
backreaction problems and remains robust under smooth deformations of
the coupling profile. Beyond magnetogenesis, the dark photon may
itself constitute a viable dark matter candidate and contribute
non-trivially to the late-time dynamics of the dark Universe.


\section{The model and dynamics}

We augment the field content minimally by positing a massless dark
photon $C_\mu$ in addition to the usual photon $A_\mu$ (equivalently,
the hypercharge field in the Standard Model). Defining the
corresponding field strength tensors by $C_{\mu\nu}$ and $A_{\mu\nu}$,
and denoting the usual electromagnetic current by $j_\mu$, the action
for the gauge fields is given by

\begin{eqnarray}\label{eq:Action-Our-work}
	S &=& -\frac{1}{4}\int {\rm d}^{4}x\, \sqrt{-\mathfrak{g}}\,
        \left[A_{\mu\nu}A^{\mu\nu} + f^2(\phi) C_{\mu \nu} C^{\mu \nu}
          \right.\nonumber \\ && \left. \hspace{60pt} + g(\phi)
          C_{\mu \nu} A^{\mu \nu} + j^\mu A_{\mu} \right],
\end{eqnarray}
where the inflaton $\phi$ is a homogeneous real scalar with the
functions $f(\phi)$ and $g(\phi)$ encapsulating its interactions with
the gauge sector. Assuming the FLRW metric with the scale factor
$a(\eta)$, $\eta$ being the conformal time, the electric and the
magnetic fields are expressed as

\begin{equation}\label{eq:define E B}
  E_{i}=- a^{-1}A_{i}^{\prime} , \qquad
  B_{i}= a^{-1}\epsilon_{ijk}\partial_{j}A_{k},
\end{equation}
and similarly for the $C$-field. The analysis is simplified in the
Coulomb gauge, {\em viz.}  $A_{0}=C_0 = 0$ and $\partial_{i}A^{i}=
\pa_i{C^i} = 0$. In the early Universe, $j^{\mu} \approx 0$ and the
gauge field equations of motion are

\begin{eqnarray}\label{eq:eqA}
  && 4h^2(A_i^{\prime \prime} - \nabla^2 A_i) =  g g^\prime A_i^\prime
  + 2 f \left(g f^\prime - 2 f g^\prime\right) C_i^\prime  \\
	\label{eq:eqB}
	&& 4 h^2(C_i^{\prime \prime} - \nabla^2 C_i) = (g g^\prime - 8 f f^\prime) 
              C_i^\prime - 2 g^\prime A_i^\prime.
\end{eqnarray}
with $h^2 \equiv f^2 - \frac{1}{4} g^2$. The cross-coupling 
renders the equations analytically intractable as also prevents a
straightforward quantization with Bunch-Davies initial
conditions. It is therefore useful to perform a field redefinition
\begin{eqnarray}\label{eq:can}
  A_i = \mathcal{A}_i - (g/2h)\, \mathcal{C}_i, \quad
  C_i = h^{-1}\, \mathcal{C}_i
\end{eqnarray}
that partially diagonalizes the action, {\em viz.},
\begin{equation}
    \label{eq:dec-Lagrangian}
\barr{rcl}
S &=& \dis \frac{1}{2}\int {\rm d}^4 {\bf x}\Big[ \mathcal{A}_i^\prime{}^2 +  \mathcal{C} i^\prime{}^2 - \left(\partial_{i}\mathcal{A}^i\right)^2 -  \left(\partial_{i}\mathcal{C}^i\right)^2 \\[1.5ex]
  && \dis \hspace*{2em}+ \frac{h^{\prime \prime}}{ h} \mathcal{C}_i^2
          + k_\ast^2 \mathcal{C}_i^2  -  k_\ast^\prime \mathcal{A}_i \mathcal{C}^i  - 2 a^2 j_i \mathcal{A}_i\Big] \ , \\[1ex]
k_\ast &\equiv & (g^\prime/2 h) \ .
\earr
\end{equation}
This field redefinition is essentially unique (up to trivial
rescalings and constant rotations) in removing the kinetic mixing
while rendering both fields canonically normalized.  Any alternative
transformation achieving canonical normalization can differ only by a
rotation of the already canonical fields and thus cannot eliminate the
residual mixing term; the $\mathcal{A}_i \mathcal{C}^i$ interaction is
unavoidable and reflects the time-dependent coupling
structure. Importantly, Eq. \eqref{eq:dec-Lagrangian} is already
sufficient for consistent canonical quantization with Bunch–Davies
initial conditions, as the remaining mixing is non-derivative;
a full diagonalization at all times is neither necessary nor
generically possible. However, as the mixing term still renders the
analysis cumbersome, a considerable simplification may be achieved by
assuming, at early times, 
$k_\ast$ is constant 
or, at most, only very slowly varying with time. This leads to
\begin{equation}
\barr{rcl}
  \mathcal{A}_i^{\prime\prime} -  \nabla^2\mathcal{A}_i & = & 0 \\
 \mathcal{C}_i^{\prime\prime} - \left(\nabla^2 + k_\ast^2
        + h^{-1} \,  h^{\prime \prime} \right)\mathcal{C}_i &=& 0
\earr
   \label{eq:eqAC}        
\end{equation}
Once the equations are decoupled\footnote{It suffices if the fields
remain effectively decoupled at an early epoch so that the
Bunch-Davies conditions can be applied.  A complete decoupling is
neither necessary, nor is it feasible in an interacting theory. The
negative conclusions of ref.~\cite{Ferreira:2014mwa} largely stemmed
from such an effort.}, canonical quantization proceeds in the standard manner,
leading to the mode expansion
\begin{eqnarray}\label{eq:FourierAB}
  \hat{\mathcal{A}}_{i}\left(\eta, \x \right)
  =\sum_{\sigma}\int\frac{\d^{3}\k}{\left(2\pi\right)^{3/2}}
  \left[\epsilon_{i \sigma}^{\mathcal{A}}(\k)\hat{a}^\sigma_{\k}\,
    \mathcal{A}_{\k}\left(\eta\right)e^{i\,\k.\x}+ h.c \right]
\end{eqnarray}
where $\epsilon_{i\,\sigma}^{\mathcal{A}}(\k)$ denotes the normalized
polarization vector for the spin polarization $\sigma$. An analogous
expression obtains for $\hat{\mathcal{C}}_{i}\left(\eta, \x \right)$.
The respective annihilation ($\hat{a}^{
  \sigma}_{\k},\hat{c}^{\sigma}_{\k}$) and creation operators
$(\hat{a}^{\sigma \dagger}_{\k}, \hat{c}^{\sigma \dagger}_{\k})$
decouple and obey standard commutation relations\footnote{Since the
interaction becomes active only over a finite interval, the visible
and dark sectors admit well-defined asymptotic free modes before and
after the transition. The creation and annihilation operators are,
therefore, well-defined with respect to the asymptotic free vacuum prior to
the onset of the interaction.
Physically, the transient interaction induces a
temporary transfer of power between the fields, rather than a
persistent mixing requiring complete diagonalization of the system at
all times.}.

\subsection{Energy spectra}

At this stage, one can explicitly compute the energy density contained
in the decoupled fields ${\cal A}_\mu$ and ${\cal C}_\mu$. However, it
is physically more meaningful to transform back to the original fields
$A_\mu$ and $C_\mu$. Using the the subscripts $E$($M$) to refer to
electric(magnetic) contributions respectively, we then have
\begin{eqnarray}
	\label{eq:mag-spec}
	&&\mathcal{P}^A_{\rm M} = \frac{1}{2 \pi^2}\frac{k^5}{a^4} \left(\left|\mathcal{A}_{\bf k}\right|^2 + \frac{g^2}{4 h^2} \left|\mathcal{C}_{\bf k}\right|^2\right), 
	\\
  	&&\mathcal{P}^A_{\rm E} = \frac{1}{2 \pi^2} \frac{k^3}{a^4} \left(\left|\mathcal{A}^\prime_{\bf k}\right|^2 + \left|\left(\frac{g^2}{2h} \mathcal{C}_{\bf k} \right)^\prime\right|^2\right), \\
  && \mathcal{P}^C_{\rm M} = \left(1 + \frac{g^2}{4 h^2}\right)\frac{k^5}{2 \pi^2a^4} \left|\mathcal{C}_{\bf k}\right|^2,\\
	&& \mathcal{P}^C_{\rm E} = \left(1 + \frac{g^2}{4 h^2}\right) \frac{k^3}{2 \pi^2a^4} \left|\mathcal{C}^\prime_{\bf k} - \frac{h^\prime}{h}\mathcal{C}_{\bf k}\right|^2,\\
  &&  \mathcal{P}^{\rm mix}_{\rm M} = \frac{g^2}{2h^2} \frac{k^5}{4 \pi^2 a^4}
  \left|\mathcal{C}_{\bf k}\right|^2,\\
	&&  \mathcal{P}^{\rm mix}_{\rm E} = \frac{g}{2 \pi^2}\frac{k^3}{a^4} \left|\left(\frac{g^2 \mathcal{C}_{\bf k}}{2 h}\right)^\prime \left(\frac{\mathcal{C}_{\bf k}}{h}\right)^\prime\right|.\label{eq:mix_E_spec}
\end{eqnarray}
where the “mix” terms encode energy transfer between the visible and
dark sectors due to the original kinetic mixing. We focus, henceforth, on $\mathcal{P}^A_{\rm M}$ which determines the
observable magnetic field at the end of inflation, {\em viz.}
\begin{equation}\label{eq:mag end inf A}
B_{\rm end}(k, a) = \sqrt{\left.\mathcal{P}^A_{\rm M}(k, a)\right|_{\rm end}}.
\end{equation}

\subsection{Revisiting inflationary solutions and avoiding backreaction}\label{sec:revisit-inflation}

To determine the evolution of the mode functions $\mathcal{A}_{\bf k}$
and $\mathcal{C}_{\bf k}$, and hence the magnetic fields generated
during inflation, it is necessary to specify the functional forms of
the coupling functions $f(\eta)$ and $g(\eta)$, or equivalently
$h(\eta)$. Before doing so, it is useful to clearly articulate the
theoretical requirements that any viable inflationary magnetogenesis
scenario must satisfy:

\begin{itemize}
\item[(i)] the generated magnetic field spectrum should be
  significantly  large {\bf without backreaction} on
  cosmological scales;
\item[(ii)] the theory must remain {\bf weakly coupled} throughout
  inflation so that perturbation theory remains valid and loop
  corrections do not dominate.
\end{itemize}
While each of these conditions can be satisfied individually,
realizing  both simultaneously over the entire
inflationary epoch is highly nontrivial and lies at the heart of the
strong-coupling and backreaction problems.

This is best understood by considering Eqs. (\ref{eq:eqAC}). While
${\cal A}_i$ would not be expected to amplify, ${\cal C}_i$ may be,
depending on the form of $h(\eta)$. As refs. \cite{Demozzi2009,
  Nandi2021InflMag} show, the amplification would be maximal for
$h(\eta) = h_0(\eta_{\rm end}/\eta)^2$. The requirement of decoupling
($k_\ast'\approx 0$) would, then, stipulate $g(\eta) = g_1 + g_2
\left(\eta_{\rm end}/\eta\right)$. However, note that the effective
gauge coupling associated with the dark photon scales as $(g/2 h)$
and, thus, for $\eta/\eta_{\rm end} \gg 1$, the above solution implies
that the coupling grows without bound in the far past, bringing us
back to the strong coupling problem.
 
To circumvent this, we posit, instead, that the interaction $g(\eta)$
switches on only after a time $\eta_\ast$, {\em viz.},
\begin{equation}
	g(\eta) =
	\begin{cases}
		0, & \eta < \eta_\ast, \\
		g_0, & \eta \ge \eta_\ast,
	\end{cases}
\end{equation}
with $g_0$ a constant. This choice ensures that $g'/2h = 0$ everywhere
except during the transition, so that the equations of motion remain
decoupled for almost the entire evolution. The effective coupling is
nonzero only in the interval $\eta_\ast \le \eta \le \eta_{\rm end}$
and attains its maximum value $g_0/2h(\eta_\ast)$.  Requiring
perturbativity at all times requires that this ratio is smaller than
unity, which, together with the normalization condition $f(\eta_{\rm
  end}) = 1$, fixes
\begin{equation}
h_0 = \left[1 + (\eta_{\rm end}/\eta_\ast)^2\right]^{-1/2}, \quad
g_0 = 2(\eta_{\rm end}/\eta_\ast)^2 \, h_0 
\end{equation}
Thus, for $\eta_* \gg \eta_{\rm end}$,
while $h_0 \sim \mathcal{O}(1)$, the kinetic mixing parameter satisfies
$g_0 \ll 1$.
\paragraph*{Mode evolution and magnetic-field amplitude.}
 Imposing Bunch-Davies initial conditions, the mode functions can be
 written schematically as
\begin{eqnarray}
	\mathcal{A}_{\bf k} &=& \frac{1}{\sqrt{2k}} e^{-i k \eta} + \frac{1}{k} \int_{- \infty}^{\eta} d \eta^\prime \sin [k (\eta - \eta^\prime)] S(\eta), \\
	\mathcal{C}_{\bf k} &\simeq& \frac{\sqrt{\pi}}{2}\sqrt{-\eta}\,
	H^{(1)}_{5/2}(-k\eta) ,
\end{eqnarray}
where $S(\eta) \equiv - k_\ast^\prime \mathcal{C}_{\bf k}$ encodes the
residual mixing in Eq. \eqref{eq:dec-Lagrangian}. The interaction is
non-vanishing only within a narrow interval around the transition, so
that the fields evolve independently at early times. Hence, on
super-Hubble scales, the homogeneous solutions satisfy
\begin{equation}
|\mathcal{A}_{\bf k}| \sim k^{-1},\qquad |\mathcal{C}_k|^2 \simeq k^{-1}(-k\eta)^{-4},
\end{equation}
implying $|\mathcal{C}_k| \gg |\mathcal{A}_k|$ before the transition.. During the transition
interval, the residual mixing term sources $\mathcal{A}_k$ from
$\mathcal{C}_k$, and the late-time amplitude of $\mathcal{A}_{\bf k}$
is therefore dominated by this sourced contribution. As for the
reciprocal sourcing of $\mathcal{C}_k$ by $\mathcal{A}_k$, this
contribution is parametrically suppressed and the induced correction
remains negligible. Consequently, $\mathcal{C}_k$ is well approximated
by its homogeneous solution throughout the evolution. Substituting
into Eq.~\eqref{eq:mag-spec}, the dominant contribution to the
observable magnetic field arises from the term proportional to
$|\mathcal{C}_k|^2$, and the magnetic-field power spectrum becomes
\begin{equation}
	\left.\mathcal{P}^A_{\rm M}\right|_{\rm end}
	\simeq \frac{1}{2\pi^2}\frac{g_0^2}{4h_0^2} H_{\rm end}^4,
\end{equation}
leading to
\begin{equation}
	B_{\rm end} \sim \frac{g_0}{2h_0} H_{\rm end}^2
	\simeq \left(\frac{\eta_{\rm end}}{\eta_\ast}\right)^2 H_{\rm end}^2.
\end{equation}
For $H_{\rm end}^2 \sim 10^{-10} M_{\rm Pl}^2$ and $\eta_\ast =
10\,\eta_{\rm end}$, this yields $B_{\rm end} \sim 10^{44}\,{\rm
  G}$. And, as $g_0/2h_0 \sim 10^{-2}$ at the end of inflation, the
ratio of the standard and dark magnetic fields at that stage,
is $ \sim 10^{-2}.$

\subsection{Impact of a Smooth, Transient Coupling}

Although, for the sake of a simple analytical solution, we have
modeled the interaction $g(\eta)$ using an idealized step-like
profile, the qualitative conclusions remain unchanged if the
transition is smoothed over a finite interval. In particular, we
consider a hyperbolic tangent profile,
\begin{equation}
	g(\eta) = \frac{g_0}{2}\left[1 + \tanh\!\left(\frac{N - N_\ast}{\Delta N}\right)\right],
	\label{eq:smoothg}
\end{equation}
where $N$ denotes the number of e-folds, $N_\ast$ corresponds to the
onset of the interaction, and $\Delta N$ controls the duration of the
transition. Throughout this analysis, we fix the onset of the
interaction at $\eta_\ast = 10\,\eta_{\rm end}$ which corresponds to
$N_\ast \simeq 57.7$ assuming a total duration of $N \simeq 60$
e-folds of inflation with the scale factor at the end of inflation
$a_{\rm end} = 10^{-29}$ and the Hubble parameter $H = 10^{-5} M_{\rm
  Pl}$.  This choice ensures that the transient interaction becomes
active only during the final stages of inflation, when the relevant
cosmological modes are already super-Hubble.
\begin{figure}[b]
	\centering
	\includegraphics[width=0.45\textwidth, height=0.27\textwidth]{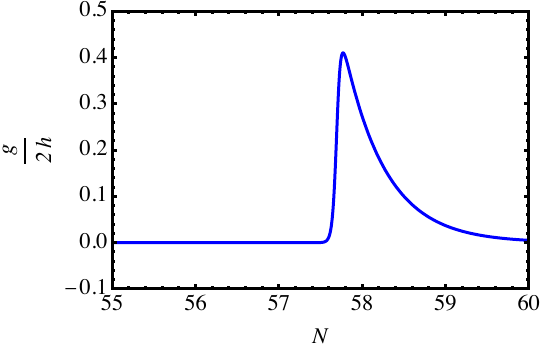}
	\caption{Time evolution of the effective coupling $g/2h$ for a
          smooth, transient interaction modeled by the hyperbolic
          tangent profile in Eq.~\eqref{eq:smoothg} with $\Delta N =
          1/20$. The vertical marker indicates the onset of the
          interaction at $N_\ast = 57.7$ (corresponding to $\eta_\ast
          = 10\,\eta_{\rm end}$).  The coupling remains perturbative
          at all times and becomes appreciable only within a narrow
          interval around the transition.}
	\label{fig:geff}
\end{figure}

To explicitly demonstrate the impact of such a smooth, transient
coupling, we numerically evolve the system using the profile in
Eq.~\eqref{eq:smoothg}. Figure~\ref{fig:geff} shows the evolution of
the effective coupling $g/2h$.  As is evident, the coupling remains
perturbative throughout the evolution, and its derivative is
appreciable only within a narrow transition interval.

\begin{figure}[t]
	\centering
	\includegraphics[width=0.45\textwidth, height=0.27\textwidth]{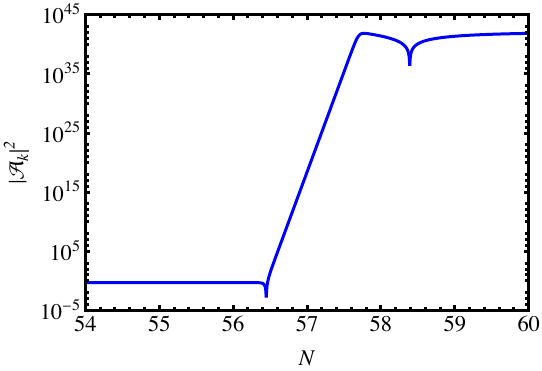}
	\includegraphics[width=0.45\textwidth, height=0.27\textwidth]{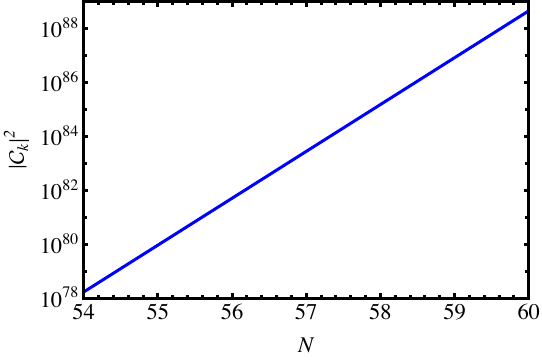}
        
	\caption{Evolution of the mode amplitudes in the presence of a
          smooth, transient coupling with $\Delta N = 1/20$ which is
          switched on at $N_\ast = 57.7$ ($\eta_\ast = 10\,\eta_{\rm
            end}$). {\em (top) $|\mathcal{A}_k|^2$ exhibits standard
            super-Hubble freezing, with only mild transient features
            during the coupling transition. {\em (bottom)} For
            $|\mathcal{C}_k|^2$}, the super-Hubble amplification is
          insensitive to the smoothing of the interaction and closely
          matches the sudden-transition limit, providing the dominant
          contribution to the magnetic field spectrum.}
	\label{fig:Amode}
\end{figure}

Having established the controlled behavior of the coupling functions,
we now examine the corresponding evolution of the mode
amplitudes. Figures~\ref{fig:Amode} display the evolution of the
canonicalized variables $\mathcal{A}_k$ and $\mathcal{C}_k$,
for $k = 1$ Mpc${}^{-1}$. While the $\mathcal{A}_{ k}$
modes do exhibit a mild response during the transition\footnote{The
sharp spikes are just artefacts of plotting $|{\mathcal A}_k|^2$near
the oscillation zeroes of ${\mathcal A}_k$ which are given in terms of
Bessel (or Hankel) functions.}  the $\mathcal{C}_{ k}$ modes remain
essentially unaffected by the smoothing of the coupling and undergo
the same super-Hubble amplification as in the sudden transition
limit. This distinction between the response of $\mathcal{A}_k$ and
$\mathcal{C}_k$ is crucial for understanding the robustness of the
mechanism.

The leading contribution to the standard magnetic field arises from
the dark photon sector through the mixing term, rather than from the
intrinsic amplification of $\mathcal{A}_{ k}$ itself.  Since transient
features induced in $\mathcal{A}_{k}$ by the smooth coupling profile
do not significantly affect the final magnetic field strength,
smoothening the interaction over a finite interval does not alter the
resulting magnetogenesis, demonstrating the robustness of the
mechanism and confirms that the analytical estimates of
Sec. \ref{sec:revisit-inflation} remain valid beyond the
sudden-transition limit. With the energy density stored in the
generated gauge fields remaining sub-dominant throughout---even in the
presence of the transient mixing term---the induced backreaction
remains sufficiently suppressed so as not to affect the inflationary
background dynamics.

\subsection{Post-inflationary evolution and the current abundance of the magnetic field}

At the end of inflation, the inflaton decays into relativistic
particles, leading to the onset of the radiation-dominated (and, then,
the matter-dominated) epoch. Assuming instantaneous reheating, the
Universe rapidly becomes highly conductive ($\sigma \gg H$).  Going
back to the cosmic time $t$, the super-Hubble equations of
motion during the radiation-dominated era are

\begin{equation}
\ddot{A}_i + \sigma \dot{A}_i \simeq 0 \ , 
\qquad 
\ddot{C}_i + H \dot{C}_i + \frac{k^2}{a^2} C_i 
\simeq \frac{g_0}{2}\sigma \dot{A}_i \ ,
\end{equation}
where $H=1/(2t)$ and $a(t)\propto t^{1/2}$. This gives
\begin{equation}
\vec A = \vec\alpha_1 +\vec\alpha_2\,e^{-\sigma t},
\end{equation}
with the constants of integration $\vec \alpha_i(\vec x)$ fixed by
matching with the inflationary solution at the reheating epoch.  The
exponential damping of the ordinary electric field rapidly renders the
source term for $\vec C$ negligible, and the subsequent evolution of
$C_i$ is essentially governed by the homogeneous equation. Retaining
the finite momentum contribution, and denoting $\kappa^2\equiv k^2
t_r/a_r^2$ (the subscript $r$ denoting the reheating epoch), we have

\begin{equation}
\vec C_k (t)
=
\vec \beta_1 \, \cos(2\kappa\sqrt{t})
+
\vec \beta_2 \sin(2\kappa\sqrt{t}),
\end{equation}
where, once again, $\vec \beta_i(\vec k)$ are fixed by matching. This
corresponds, in Fourier space, to the standard plane-wave solution of
a free massless gauge field propagating in a radiation-dominated
Universe. In other words, post-reheating the dark photon behaves as a
free oscillating radiation mode with no further amplification or
instability.  Even in the matter-dominated era, the oscillating
behaviour is maintained, albeit with a slightly different frequency.

Thus, while the ordinary electric field $E_A$ 
  decays exponentially in time, both the magnetic fields $B_A$ and $B_C$
  as well the dark electric field $E_C$ are redshifted away only
as $\sim a^{-2}.$ Consequently,  the current abundance of the
ordinary magnetic field $B_A$ is
\[
B^{\rm max}_0 \sim 10^{-14} ~\text{G},
\]
while the dark magnetic field $B_C$ has relaxed down to $\sim
10^{-12}$ G, or $\sim 100$ times higher than the standard
one\footnote{And while the naive expectation for $E_C$ is similar,
note that it suffers an additional decay on account of the Schwinger
effect, albeit at a small rate on account of the suppressed coupling
$C_\mu$ has with the charged fields.} The consequences of such
enhanced dark magnetic fields constitute a broad subject of
investigation, which may have implications for dark matter
phenomenology. This is beyond the scope of this work and we
reserve the discussion for future endeavors.

\subsection{Post-inflationary relaxation of the coupling}
Irrespective of the exact form of $g(\eta)$, as long as its
post-inflationary evolution satisfies the adiabaticity condition,
namely $\dot g \ll H g$, it would undergo a slow decay over many
Hubble times. This ensures that the coupling evolution does not excite
additional gauge modes. In this case, even a moderate value at the end
of inflation, $g_{\rm end} \sim 10^{-2}\text{--}10^{-4}$, can be
significantly suppressed during the subsequent radiation-dominated
era, reaching values as small as $g \sim 10^{-10}\text{--}10^{-15}$
without affecting the inflationary magnetogenesis mechanism. This ensures that current experimental
bounds on photon mixings are easily satisfied.

So far, we have assumed that the field $C$ is massless. While
simplicity required this to be true during the inflationary era,
subsequently the corresponding gauge symmetry could have been broken
spontaneously, thereby rendering $C$ massive and evading the bounds on
the mixing. Note, though, that the $C$-charge of the scalar
responsible would have to be tiny so as to evade bounds from
nonperturbativity. While this can be arranged and is attendant with
very interesting phenomenology, we postpone this for a future
analysis.

\section{Conclusion}

A minimal scenario incorporating a time-dependent kinetic mixing
between the photon and its dark cousin is, thus, shown to accommodate
inflationary magnetogenesis without the usual problems of back
reaction and/or loss of perturbativity.  The amplification is driven
primarily by the dark photon sector and is efficiently transferred to
the visible sector.

We further showed that the mechanism is robust under smooth
deformations of the coupling profile and remains consistent during the
post-inflationary evolution in a highly conducting plasma, leading to
present-day magnetic field strengths compatible with current
observational bounds. An intriguing open direction concerns the role
of the dark photon itself: depending on its mass generation mechanism
and subsequent evolution, it may constitute a viable dark matter
candidate and contribute non-trivially to the late-time dark Universe
through its residual magnetic fields and interactions. Understanding
the microphysical origin of the transient coupling, its possible
realization during reheating, and the phenomenological consequences
for dark matter and late-time cosmology remain important open
questions, which we leave for future endeavors.

\section*{acknowledgments}
DN acknowledges the DST (India) for support through the DST--INSPIRE
Faculty Fellowship (Grant No.~04/2020/002142).  DC acknowledges the
ANRF (India) for support through the project CRG/2023/008234 and the
IoE, University of Delhi grant IoE/2025-26/12/FRP.


\begin{thebibliography}{23}%
	\makeatletter
	\providecommand \@ifxundefined [1]{%
		\@ifx{#1\undefined}
	}%
	\providecommand \@ifnum [1]{%
		\ifnum #1\expandafter \@firstoftwo
		\else \expandafter \@secondoftwo
		\fi
	}%
	\providecommand \@ifx [1]{%
		\ifx #1\expandafter \@firstoftwo
		\else \expandafter \@secondoftwo
		\fi
	}%
	\providecommand \natexlab [1]{#1}%
	\providecommand \enquote  [1]{``#1''}%
	\providecommand \bibnamefont  [1]{#1}%
	\providecommand \bibfnamefont [1]{#1}%
	\providecommand \citenamefont [1]{#1}%
	\providecommand \href@noop [0]{\@secondoftwo}%
	\providecommand \href [0]{\begingroup \@sanitize@url \@href}%
	\providecommand \@href[1]{\@@startlink{#1}\@@href}%
	\providecommand \@@href[1]{\endgroup#1\@@endlink}%
	\providecommand \@sanitize@url [0]{\catcode `\\12\catcode `\$12\catcode
		`\&12\catcode `\#12\catcode `\^12\catcode `\_12\catcode `\%12\relax}%
	\providecommand \@@startlink[1]{}%
	\providecommand \@@endlink[0]{}%
	\providecommand \url  [0]{\begingroup\@sanitize@url \@url }%
	\providecommand \@url [1]{\endgroup\@href {#1}{\urlprefix }}%
	\providecommand \urlprefix  [0]{URL }%
	\providecommand \Eprint [0]{\href }%
	\providecommand \doibase [0]{http://dx.doi.org/}%
	\providecommand \selectlanguage [0]{\@gobble}%
	\providecommand \bibinfo  [0]{\@secondoftwo}%
	\providecommand \bibfield  [0]{\@secondoftwo}%
	\providecommand \translation [1]{[#1]}%
	\providecommand \BibitemOpen [0]{}%
	\providecommand \bibitemStop [0]{}%
	\providecommand \bibitemNoStop [0]{.\EOS\space}%
	\providecommand \EOS [0]{\spacefactor3000\relax}%
	\providecommand \BibitemShut  [1]{\csname bibitem#1\endcsname}%
	\let\auto@bib@innerbib\@empty
	\bibitem [{\citenamefont {Kandus}\ \emph {et~al.}(2011)\citenamefont {Kandus},
		\citenamefont {Kunze},\ and\ \citenamefont {Tsagas}}]{Kandus:2010nw}%
	\BibitemOpen
	\bibfield  {author} {\bibinfo {author} {\bibfnamefont {A.}~\bibnamefont
			{Kandus}}, \bibinfo {author} {\bibfnamefont {K.~E.}\ \bibnamefont {Kunze}}, \
		and\ \bibinfo {author} {\bibfnamefont {C.~G.}\ \bibnamefont {Tsagas}},\
	}\href {\doibase 10.1016/j.physrep.2011.03.001} {\bibfield  {journal}
		{\bibinfo  {journal} {Phys. Rept.}\ }\textbf {\bibinfo {volume} {505}},\
		\bibinfo {pages} {1} (\bibinfo {year} {2011})},\ \Eprint
	{http://arxiv.org/abs/1007.3891} {arXiv:1007.3891 [astro-ph.CO]} \BibitemShut
	{NoStop}%
	\bibitem [{\citenamefont {Durrer}\ and\ \citenamefont
		{Neronov}(2013)}]{Durrer:2013pga}%
	\BibitemOpen
	\bibfield  {author} {\bibinfo {author} {\bibfnamefont {R.}~\bibnamefont
			{Durrer}}\ and\ \bibinfo {author} {\bibfnamefont {A.}~\bibnamefont
			{Neronov}},\ }\href {\doibase 10.1007/s00159-013-0062-7} {\bibfield
		{journal} {\bibinfo  {journal} {Astron. Astrophys. Rev.}\ }\textbf {\bibinfo
			{volume} {21}},\ \bibinfo {pages} {62} (\bibinfo {year} {2013})},\ \Eprint
	{http://arxiv.org/abs/1303.7121} {arXiv:1303.7121 [astro-ph.CO]} \BibitemShut
	{NoStop}%
	\bibitem [{\citenamefont {Subramanian}(2016)}]{Subramanian:2015lua}%
	\BibitemOpen
	\bibfield  {author} {\bibinfo {author} {\bibfnamefont {K.}~\bibnamefont
			{Subramanian}},\ }\href {\doibase 10.1088/0034-4885/79/7/076901} {\bibfield
		{journal} {\bibinfo  {journal} {Rept. Prog. Phys.}\ }\textbf {\bibinfo
			{volume} {79}},\ \bibinfo {pages} {076901} (\bibinfo {year} {2016})},\
	\Eprint {http://arxiv.org/abs/1504.02311} {arXiv:1504.02311 [astro-ph.CO]}
	\BibitemShut {NoStop}%
	\bibitem [{\citenamefont {Vachaspati}(2021)}]{Vachaspati:2020blt}%
	\BibitemOpen
	\bibfield  {author} {\bibinfo {author} {\bibfnamefont {T.}~\bibnamefont
			{Vachaspati}},\ }\href {\doibase 10.1088/1361-6633/ac03a9} {\bibfield
		{journal} {\bibinfo  {journal} {Rept. Prog. Phys.}\ }\textbf {\bibinfo
			{volume} {84}},\ \bibinfo {pages} {074901} (\bibinfo {year} {2021})},\
	\Eprint {http://arxiv.org/abs/2010.10525} {arXiv:2010.10525 [astro-ph.CO]}
	\BibitemShut {NoStop}%
	\bibitem [{\citenamefont {Turner}\ and\ \citenamefont
		{Widrow}(1988)}]{Turner:1987bw}%
	\BibitemOpen
	\bibfield  {author} {\bibinfo {author} {\bibfnamefont {M.~S.}\ \bibnamefont
			{Turner}}\ and\ \bibinfo {author} {\bibfnamefont {L.~M.}\ \bibnamefont
			{Widrow}},\ }\href {\doibase 10.1103/PhysRevD.37.2743} {\bibfield  {journal}
		{\bibinfo  {journal} {Phys. Rev. D}\ }\textbf {\bibinfo {volume} {37}},\
		\bibinfo {pages} {2743} (\bibinfo {year} {1988})}\BibitemShut {NoStop}%
	\bibitem [{\citenamefont {Ratra}(1992)}]{Ratra:1991bn}%
	\BibitemOpen
	\bibfield  {author} {\bibinfo {author} {\bibfnamefont {B.}~\bibnamefont
			{Ratra}},\ }\href {\doibase 10.1086/186384} {\bibfield  {journal} {\bibinfo
			{journal} {Astrophys. J. Lett.}\ }\textbf {\bibinfo {volume} {391}},\
		\bibinfo {pages} {L1} (\bibinfo {year} {1992})}\BibitemShut {NoStop}%
	\bibitem [{\citenamefont {Demozzi}\ \emph {et~al.}(2009)\citenamefont
		{Demozzi}, \citenamefont {Mukhanov},\ and\ \citenamefont
		{Rubinstein}}]{Demozzi:2009fu}%
	\BibitemOpen
	\bibfield  {author} {\bibinfo {author} {\bibfnamefont {V.}~\bibnamefont
			{Demozzi}}, \bibinfo {author} {\bibfnamefont {V.}~\bibnamefont {Mukhanov}}, \
		and\ \bibinfo {author} {\bibfnamefont {H.}~\bibnamefont {Rubinstein}},\
	}\href {\doibase 10.1088/1475-7516/2009/08/025} {\bibfield  {journal}
		{\bibinfo  {journal} {JCAP}\ }\textbf {\bibinfo {volume} {08}},\ \bibinfo
		{pages} {025} (\bibinfo {year} {2009})},\ \Eprint
	{http://arxiv.org/abs/0907.1030} {arXiv:0907.1030 [astro-ph.CO]} \BibitemShut
	{NoStop}%
	\bibitem [{\citenamefont {Fujita}\ and\ \citenamefont
		{Mukohyama}(2012)}]{Fujita:2012rb}%
	\BibitemOpen
	\bibfield  {author} {\bibinfo {author} {\bibfnamefont {T.}~\bibnamefont
			{Fujita}}\ and\ \bibinfo {author} {\bibfnamefont {S.}~\bibnamefont
			{Mukohyama}},\ }\href {\doibase 10.1088/1475-7516/2012/10/034} {\bibfield
		{journal} {\bibinfo  {journal} {JCAP}\ }\textbf {\bibinfo {volume} {10}},\
		\bibinfo {pages} {034} (\bibinfo {year} {2012})},\ \Eprint
	{http://arxiv.org/abs/1205.5031} {arXiv:1205.5031 [astro-ph.CO]} \BibitemShut
	{NoStop}%
	\bibitem [{\citenamefont {Kobayashi}(2014)}]{Kobayashi:2014sga}%
	\BibitemOpen
	\bibfield  {author} {\bibinfo {author} {\bibfnamefont {T.}~\bibnamefont
			{Kobayashi}},\ }\href {\doibase 10.1088/1475-7516/2014/05/040} {\bibfield
		{journal} {\bibinfo  {journal} {JCAP}\ }\textbf {\bibinfo {volume} {05}},\
		\bibinfo {pages} {040} (\bibinfo {year} {2014})},\ \Eprint
	{http://arxiv.org/abs/1403.5168} {arXiv:1403.5168 [astro-ph.CO]} \BibitemShut
	{NoStop}%
	\bibitem [{\citenamefont {Anber}\ and\ \citenamefont
		{Sorbo}(2006)}]{Anber:2006xt}%
	\BibitemOpen
	\bibfield  {author} {\bibinfo {author} {\bibfnamefont {M.~M.}\ \bibnamefont
			{Anber}}\ and\ \bibinfo {author} {\bibfnamefont {L.}~\bibnamefont {Sorbo}},\
	}\href {\doibase 10.1088/1475-7516/2006/10/018} {\bibfield  {journal}
		{\bibinfo  {journal} {JCAP}\ }\textbf {\bibinfo {volume} {10}},\ \bibinfo
		{pages} {018} (\bibinfo {year} {2006})},\ \Eprint
	{http://arxiv.org/abs/astro-ph/0606534} {arXiv:astro-ph/0606534} \BibitemShut
	{NoStop}%
	\bibitem [{\citenamefont {Ferreira}\ \emph {et~al.}(2013)\citenamefont
		{Ferreira}, \citenamefont {Jain},\ and\ \citenamefont
		{Sloth}}]{Ferreira:2013sqa}%
	\BibitemOpen
	\bibfield  {author} {\bibinfo {author} {\bibfnamefont {R.~J.~Z.}\
			\bibnamefont {Ferreira}}, \bibinfo {author} {\bibfnamefont {R.~K.}\
			\bibnamefont {Jain}}, \ and\ \bibinfo {author} {\bibfnamefont {M.~S.}\
			\bibnamefont {Sloth}},\ }\href {\doibase 10.1088/1475-7516/2013/10/004}
	{\bibfield  {journal} {\bibinfo  {journal} {JCAP}\ }\textbf {\bibinfo
			{volume} {10}},\ \bibinfo {pages} {004} (\bibinfo {year} {2013})},\ \Eprint
	{http://arxiv.org/abs/1305.7151} {arXiv:1305.7151 [astro-ph.CO]} \BibitemShut
	{NoStop}%
	\bibitem [{\citenamefont {Nandi}(2021)}]{Nandi:2021lpf}%
	\BibitemOpen
	\bibfield  {author} {\bibinfo {author} {\bibfnamefont {D.}~\bibnamefont
			{Nandi}},\ }\href {\doibase 10.1088/1475-7516/2021/08/039} {\bibfield
		{journal} {\bibinfo  {journal} {JCAP}\ }\textbf {\bibinfo {volume} {08}},\
		\bibinfo {pages} {039} (\bibinfo {year} {2021})},\ \Eprint
	{http://arxiv.org/abs/2103.03159} {arXiv:2103.03159 [astro-ph.CO]}
	\BibitemShut {NoStop}%
	\bibitem [{\citenamefont {Nandi}\ and\ \citenamefont
		{Shankaranarayanan}(2018)}]{Nandi:2017ajk}%
	\BibitemOpen
	\bibfield  {author} {\bibinfo {author} {\bibfnamefont {D.}~\bibnamefont
			{Nandi}}\ and\ \bibinfo {author} {\bibfnamefont {S.}~\bibnamefont
			{Shankaranarayanan}},\ }\href {\doibase 10.1088/1475-7516/2018/01/039}
	{\bibfield  {journal} {\bibinfo  {journal} {JCAP}\ }\textbf {\bibinfo
			{volume} {01}},\ \bibinfo {pages} {039} (\bibinfo {year} {2018})},\ \Eprint
	{http://arxiv.org/abs/1704.06897} {arXiv:1704.06897 [astro-ph.CO]}
	\BibitemShut {NoStop}%
	\bibitem [{\citenamefont {Giovannini}(2000)}]{Giovannini:1999wv}%
	\BibitemOpen
	\bibfield  {author} {\bibinfo {author} {\bibfnamefont {M.}~\bibnamefont
			{Giovannini}},\ }\href {\doibase 10.1103/PhysRevD.61.063004} {\bibfield
		{journal} {\bibinfo  {journal} {Phys. Rev. D}\ }\textbf {\bibinfo {volume}
			{61}},\ \bibinfo {pages} {063004} (\bibinfo {year} {2000})},\ \Eprint
	{http://arxiv.org/abs/hep-ph/9905358} {arXiv:hep-ph/9905358} \BibitemShut
	{NoStop}%
	\bibitem [{\citenamefont {Beltran~Jimenez}\ and\ \citenamefont
		{Maroto}(2011)}]{BeltranJimenez:2010brg}%
	\BibitemOpen
	\bibfield  {author} {\bibinfo {author} {\bibfnamefont {J.}~\bibnamefont
			{Beltran~Jimenez}}\ and\ \bibinfo {author} {\bibfnamefont {A.~L.}\
			\bibnamefont {Maroto}},\ }\href {\doibase 10.1103/PhysRevD.83.023514}
	{\bibfield  {journal} {\bibinfo  {journal} {Phys. Rev. D}\ }\textbf {\bibinfo
			{volume} {83}},\ \bibinfo {pages} {023514} (\bibinfo {year} {2011})},\
	\Eprint {http://arxiv.org/abs/1010.3960} {arXiv:1010.3960 [astro-ph.CO]}
	\BibitemShut {NoStop}%
	\bibitem [{\citenamefont {Holdom}(1986)}]{Holdom:1985ag}%
	\BibitemOpen
	\bibfield  {author} {\bibinfo {author} {\bibfnamefont {B.}~\bibnamefont
			{Holdom}},\ }\href {\doibase 10.1016/0370-2693(86)91377-8} {\bibfield
		{journal} {\bibinfo  {journal} {Phys. Lett. B}\ }\textbf {\bibinfo {volume}
			{166}},\ \bibinfo {pages} {196} (\bibinfo {year} {1986})}\BibitemShut
	{NoStop}%
	\bibitem [{\citenamefont {Redondo}\ and\ \citenamefont
		{Ringwald}(2011)}]{Redondo:2010dp}%
	\BibitemOpen
	\bibfield  {author} {\bibinfo {author} {\bibfnamefont {J.}~\bibnamefont
			{Redondo}}\ and\ \bibinfo {author} {\bibfnamefont {A.}~\bibnamefont
			{Ringwald}},\ }\href {\doibase 10.1080/00107514.2011.563516} {\bibfield
		{journal} {\bibinfo  {journal} {Contemp. Phys.}\ }\textbf {\bibinfo {volume}
			{52}},\ \bibinfo {pages} {211} (\bibinfo {year} {2011})},\ \Eprint
	{http://arxiv.org/abs/1011.3741} {arXiv:1011.3741 [hep-ph]} \BibitemShut
	{NoStop}%
	\bibitem [{\citenamefont {Arias}\ \emph {et~al.}(2012)\citenamefont {Arias},
		\citenamefont {Cadamuro}, \citenamefont {Goodsell}, \citenamefont {Jaeckel},
		\citenamefont {Redondo},\ and\ \citenamefont {Ringwald}}]{Arias:2012az}%
	\BibitemOpen
	\bibfield  {author} {\bibinfo {author} {\bibfnamefont {P.}~\bibnamefont
			{Arias}}, \bibinfo {author} {\bibfnamefont {D.}~\bibnamefont {Cadamuro}},
		\bibinfo {author} {\bibfnamefont {M.}~\bibnamefont {Goodsell}}, \bibinfo
		{author} {\bibfnamefont {J.}~\bibnamefont {Jaeckel}}, \bibinfo {author}
		{\bibfnamefont {J.}~\bibnamefont {Redondo}}, \ and\ \bibinfo {author}
		{\bibfnamefont {A.}~\bibnamefont {Ringwald}},\ }\href {\doibase
		10.1088/1475-7516/2012/06/013} {\bibfield  {journal} {\bibinfo  {journal}
			{JCAP}\ }\textbf {\bibinfo {volume} {06}},\ \bibinfo {pages} {013} (\bibinfo
		{year} {2012})},\ \Eprint {http://arxiv.org/abs/1201.5902} {arXiv:1201.5902
		[hep-ph]} \BibitemShut {NoStop}%
	\bibitem [{\citenamefont {Ferreira}\ and\ \citenamefont
		{Ganc}(2015)}]{Ferreira:2014mwa}%
	\BibitemOpen
	\bibfield  {author} {\bibinfo {author} {\bibfnamefont {R.~Z.}\ \bibnamefont
			{Ferreira}}\ and\ \bibinfo {author} {\bibfnamefont {J.}~\bibnamefont
			{Ganc}},\ }\href {\doibase 10.1088/1475-7516/2015/04/029} {\bibfield
		{journal} {\bibinfo  {journal} {JCAP}\ }\textbf {\bibinfo {volume} {04}},\
		\bibinfo {pages} {029} (\bibinfo {year} {2015})},\ \Eprint
	{http://arxiv.org/abs/1411.5362} {arXiv:1411.5362 [astro-ph.CO]} \BibitemShut
	{NoStop}%
	\bibitem [{Note1()}]{Note1}%
	\BibitemOpen
	\bibinfo {note} {It suffices if the fields remain effectively decoupled at an
		early epoch so that the Bunch-Davies conditions can be applied. A complete
		decoupling is neither necessary, nor is it feasible in an interacting theory.
		The negative conclusions of ref.~\cite {Ferreira:2014mwa} largely stemmed
		from such an effort.}\BibitemShut {Stop}%
	\bibitem [{Note2()}]{Note2}%
	\BibitemOpen
	\bibinfo {note} {Since the interaction becomes active only over a finite
		interval, the visible and dark sectors admit well-defined asymptotic free
		modes before and after the transition. The creation and annihilation
		operators are, therefore, well-defined with respect to the asymptotic free
		vacuum prior to the onset of the interaction. Physically, the transient
		interaction induces a temporary transfer of power between the fields, rather
		than a persistent mixing requiring complete diagonalization of the system at
		all times.}\BibitemShut {Stop}%
	\bibitem [{Note3()}]{Note3}%
	\BibitemOpen
	\bibinfo {note} {The sharp spikes are just artefacts of plotting $|{\protect
			\mathcal A}_k|^2$near the oscillation zeroes of ${\protect \mathcal A}_k$
		which are given in terms of Bessel (or Hankel) functions.}\BibitemShut
	{Stop}%
	\bibitem [{Note4()}]{Note4}%
	\BibitemOpen
	\bibinfo {note} {And while the naive expectation for $E_C$ is similar, note
		that it suffers an additional decay on account of the Schwinger effect,
		albeit at a small rate on account of the suppressed coupling $C_\mu $ has
		with the charged fields.}\BibitemShut {Stop}%
\end{thebibliography}

%

\end{document}